\documentclass[twocolumn,showpacs,preprintnumbers,amsmath,amssymb, superscriptaddress, prl]{revtex4}

\usepackage{graphicx}   
\usepackage{dcolumn}    
\usepackage{bm}         

\begin{document}


\title{A High Flux Source of Cold Rubidium}
%

\author{Christopher Slowe}
 \email{slowe@physics.harvard.edu}
\affiliation{Department of Physics and Division of Engineering and
Applied Sciences, Harvard University, Cambridge, Massachusetts 02138}%

\author{Laurent Vernac}%
\affiliation{Department of Physics and Division of Engineering and
Applied Sciences, Harvard University, Cambridge, Massachusetts 02138}%
\affiliation{Laboratoire de Physique des Laser, Institut
Galil\'ee, Universit\'e Paris 13, Avenue Jean Baptiste Cl\'ement
93430 Villetaneuse France}

\author{Lene Vestergaard Hau}%
\affiliation{Department of Physics and Division of Engineering and
Applied Sciences, Harvard University, Cambridge, Massachusetts 02138}%

\date{\today}

\begin{abstract}
We report the production of a continuous, slow, and cold beam of
${}^{87}\mathrm{Rb}$ atoms with an unprecedented flux of
$3.2\times 10^{12}$ atoms/s and a temperature of a few
milliKelvin.  Hot atoms are emitted from a Rb candlestick atomic
beam source and transversely cooled and collimated by a 20 cm long
atomic collimator section, augmenting overall beam flux by a
factor of 50. The atomic beam is then decelerated and
longitudinally cooled by Zeeman slowing.
\end{abstract}

\maketitle

Cold and slow, high-flux atomic beams are essential for high
signal to noise measurements in atom interferometry experiments
\cite{AtomInterferometer}, for rapid loading of magneto-optical
traps (MOTs) for production of Bose-Einstein condensates, and for
the potential creation of a CW atom laser.  A variety of methods
have been employed over the years to create cold, intense atomic
beams. Low velocity intense sources (LVIS) \cite{LVIS}, Axicons
\cite{Axicon}, and 2D MOTs \cite{AtomFunnel, 2DplusMOT,
DalibardSource} employ background vapor loading into a MOT.  They
are typically limited in total flux by the loading rate of the MOT
to a few times $10^9$ atoms/s, though they may be pushed as high
as to $6\times 10^{10}$ atoms/s, as in \cite{twoDRbMot} (but with
higher longitudinal temperature).

An alternative approach involves the use of a hot effusive source
of atoms, with a much higher vapor pressure and followed by a
deceleration and cooling stage. White light slowing has been used
to produce a beam with up to $2\times 10^{10}$ Cs atoms/s
\cite{HighFluxCsBeam}. The Zeeman slower \cite{OriginalZeeman}
requires less laser power.  Using this technique, the authors in
\cite{DalibardBeamPaper} have been able to produce a beam of Rb
with an integrated flux of $2\times 10^{11}$ atoms/s. However,
both are limited by the transverse velocity distribution of the
original atomic source and can thus be made significantly more
efficient by adding some form of transverse beam collimation and
cooling.

\begin{figure*}
\includegraphics{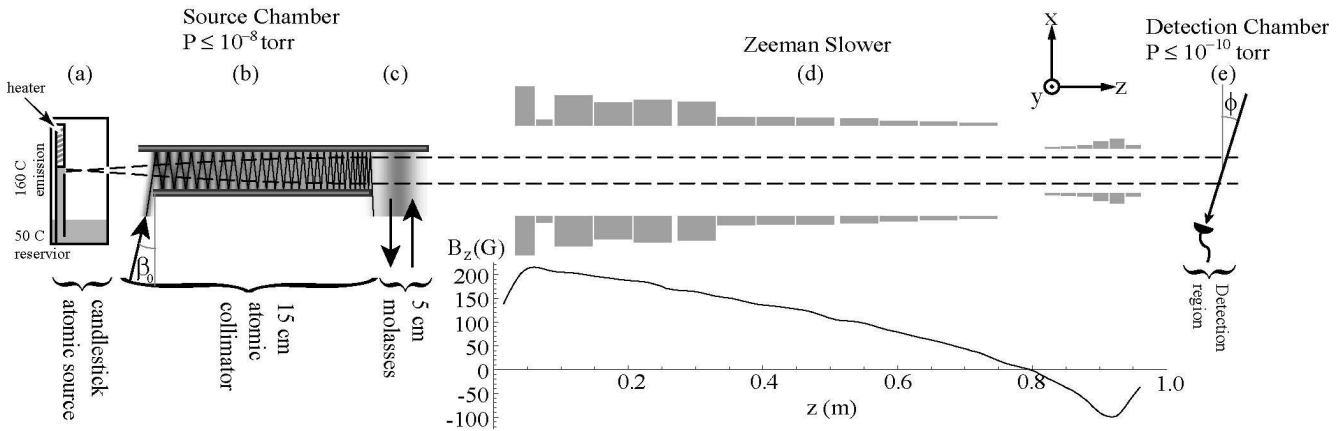}
\caption{\label{fig:setup} {\bf Experimental set-up}.  An atomic
beam is produced via a candlestick atomic beam source (a) with a
divergence angle of 1/10 radian.  The atoms in the beam then pass
through a 20 cm long transverse collimation (b) and cooling (c)
stage and are subsequently longitudinally decelerated and cooled
by means of a 1 m long Zeeman slower (d), which is designed in the
zero-crossing configuration shown. Measurement of the atomic beam
is performed via an absorption measurement 10 cm beyond the end of
the Zeeman slower (e).}
\end{figure*}

In this Letter, we report the production of a slow and cold (both
transversely and longitudinally) atomic beam with a total flux of
$3.2\times 10^{12}$ ($1.7\times 10^{12}$) atoms/s at 116 (45) m/s
and a peak intensity of $7.5\times 10^{11}$ ($3\times 10^{11}$)
atoms/s/$\mathrm{cm}^2$.  A schematic representation of the
apparatus used for production of the atomic beam is shown in
Fig.\,\ref{fig:setup}.  The set-up consists of a two-stage vacuum
chamber, with the source chamber separated from the detection
chamber by a one-meter Zeeman slower tube.

Our source of thermal atoms originates from a Rb candlestick
atomic beam source (Fig.\,\ref{fig:setup}(a)) \cite{candlestickNa,
candlestickRb}, which provides an intense, well collimated beam.
This source has a low temperature shield around a localized hot
emission point kept at $160^\circ$C, corresponding to a 7 mtorr
vapor pressure. The 2 mm emission hole lies 2 cm behind a 2 mm
collimation hole, resulting in an emitted beam with a total
divergence of $\pi/100$ sr. At this temperature, the resulting
atomic beam has a mean velocity of 400 m/s, and a total emitted
flux of $1.2\times 10^{14}$ atoms/s; $25\%$ of these emitted atoms
are ${}^{87}\mathrm{Rb}$ of which $36\%$ are within the designed
capture velocity of the Zeeman slower (320 m/s), for a total
available flux of $1.1 \times 10^{13}$ atoms/s. Rubidium at the
emission hole is replenished by capillary action via a gold-coated
stainless steel mesh wick which draws from a reservoir of liquid
rubidium (kept at $50^\circ$C just above the Rb melting point of
$40^\circ$C). The pressure of the source vacuum chamber can be
kept below $10^{-8}$ torr, with the aid of a $-45^\circ$C cold
finger. Moreover, by lining the inside of the reservoir with the
same wicking mesh, uncollimated Rb is recycled, minimizing the
need for reloading and source maintenance.

To optimize coupling of atoms from the source into the Zeeman
slower region, we employ an atomic collimator, with a large
transverse capture velocity which acts to uniformly decelerate the
transverse velocity of the atoms over its entire length
\cite{OriginalMirrorTubeIdea}.  As shown schematically in
Fig.\,\ref{fig:setup}(b), the first 15 cm of the transverse
collimation is accomplished by two pairs of nearly plane parallel
$6\times1$" mirrors (one pair for each transverse axis), with a
$2\times 1$ cm laser beam for each axis coupled in at one end of
the structure at a small angle, $\beta_0$, relative to the
mirror's normal \cite{OriginalMirrorTubeIdea,
OriginalMirrorTubeImplementation}. By an optimal choice of laser
beam and mirror geometry, we greatly increase the capture velocity
of the collimator relative to a comparable length of 2D molasses.
Here, the mirrors in each pair are at a small angle $\alpha$
relative to each other. Thus, upon the $n$th reflection, the laser
beam angle $\beta_n$ will be
\begin{equation}
    \label{eqn:beta}
    \beta_n = \beta_0 -  n \alpha.
\end{equation}
By design, as the atomic beam propagates down the collimator, it
becomes more and more collimated due to the radiation pressure
from the laser beams which become more and more normal to the
mirrors according to (\ref{eqn:beta}). In this way, it is possible
for the laser beams to remain orthogonal to the trajectory of the
atoms, which are then uniformly accelerated radially with the
maximum spontaneous force, $F_{spont}^{max}$, and follow an
approximately circular arc.

One can readily show that with flat mirrors it is impossible to
make the wavefronts follow a perfectly circular arc (as shown by
the dashed line in Fig.\,\ref{fig:setup}(b)).  We solve for
$\beta$ as a function of distance to arrive at
\begin{equation}
    \label{eqn:betaz}
    \beta(z) = \frac{1}{2}\left[\alpha +
    \sqrt{(\beta_0-\alpha)^2-\frac{4\alpha}{D}z}\right],
\end{equation}
where $D$ is the mirror spacing.  If we want the final angle to
approach zero, as it should for a perfectly collimated atomic
beam, $\beta$ should decrease linearly rather than as $\sqrt{z}$
as given by (\ref{eqn:betaz}). Although the laser rays will not
follow the ideal trajectory as a function of $z$, the increasingly
steep angle will cause the density of the rays  to increase at the
end, compensating for the intensity loss from repeated
reflections.

As a final consideration, note that we still require the laser
beam to exit the collimator; if the final angle were normal to the
mirror, the beam would simply couple back down the collimator and
undo the collimation. To avoid this, we couple the laser beam
non-orthogonally to the atomic velocity and correct for the
resulting longitudinal Doppler shift by blue detuning the laser by
$+3 \mathrm{MHz}$ ($0.5\Gamma$, where $\Gamma$ is the natural
linewidth).

Since the collimation laser is tuned to near resonance, we
sacrifice final transverse temperature for maximum collimation.
Next, the 5 cm region of 2D molasses (Fig.\,\ref{fig:setup}(c))
rectifies this by transversely cooling the atomic beam, as well as
serving as a tool for the overall alignment of the collimation
structure with respect to the Zeeman slower tube (which defines an
aperture of $10^{-3}$ sr).

With the atomic beam collimated, transversely cooled, and
efficiently coupled out of the source chamber, the longitudinal
beam slowing and cooling are accomplished with a one meter long
Zeeman slower (Fig.\,\ref{fig:setup}(d)). The slower design is of
the zero-field-crossing type with a final compensating coil
intended to leave very little fringe field in the experimental
chamber. By design, the last six coils (which set the ``negative''
part of the magnetic field profile) can be driven independently of
the first thirteen coils to allow for adjustment of the final
velocity.

We first directly demonstrate the effectiveness of the atomic beam
collimation stage by an absorption measurement at the end of the
Zeeman slower, but without the magnetic field or Zeeman slowing
beam present. By measuring the transmission coefficient of a probe
laser through the atomic beam, we may extract the optical density
(OD) of the atomic beam. Further, by sweeping the detuning,
$\Delta$, of the probe laser, we extract the atomic beam's
velocity distribution projected onto the probe's axis (see
Fig.\,\ref{fig:setup}(e)).

\begin{figure}
\includegraphics{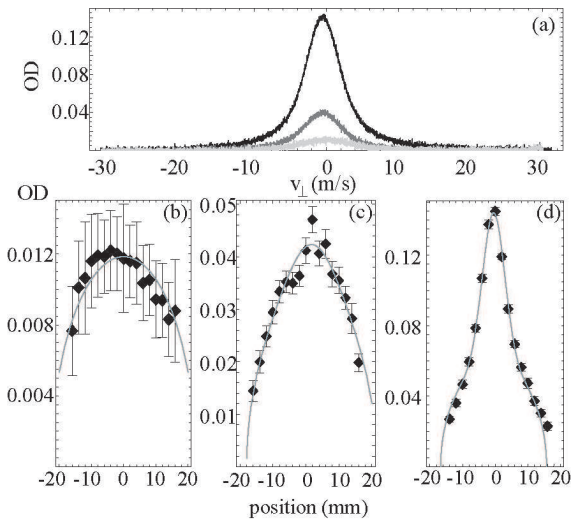}
\caption{\label{fig:normalabs} {\bf Measurement of the atomic
beam's transverse velocity and spatial distributions.} (a) OD of
the bare atomic beam (light grey) with molasses (dark grey) and
with the full 20 cm of transverse collimation and cooling (black).
(b)-(d) Transverse mode structure of the atomic beam obtained by
scanning the position of the probe beam along the direction normal
to the Zeeman slower axis and the probe beam direction.  This is
done for the (b) uncollimated atomic beam, (c) with molasses, and
(d) with the full 20 cm of transverse cooling and collimation.}
\end{figure}

Data resulting from the above method are shown in
Fig.\,\ref{fig:normalabs}(a). In this case, the probe is normal
($\phi=0^\circ$) to the axis of the Zeeman slower, so that we may
discern the transverse velocity spread.  By fitting the frequency
profile of the optical density to a Voigt distribution,
\begin{equation}
\label{eqn:voigt}
    OD(\Delta) \propto \int n(v_\perp) \sigma(\Delta-k v_\perp) \mathrm{d}v_\perp,
\end{equation}
we may extract the velocity spread of the density $n(v_\perp)$,
knowing the Lorentzian form of the cross-section $\sigma(\Delta)$
(where $v_\perp$ is the transverse velocity, and $k$ the
wavenumber). We find that the uncollimated emission from our
source has an RMS spread of 4 m/s, which is simply a consequence
of geometry (an atomic beam with a mean longitudinal velocity of
around 400 m/s apertured to a half angle of 1/100th of a radian).
Addition of the 2D molasses decreases this spread to 1.7 m/s,
while the application of the full transverse collimation and
cooling leads to a spread of 1 m/s.

As an extension of this method, in Fig.\,\ref{fig:normalabs}(b-d)
we translate the probe laser along $y$ and measure the peak OD at
each point to extract the transverse mode structure of the atomic
beam. To account for the structure, we assume that the normalized
transverse density profile, $g(x,y)$, of the atomic beam takes the
form:
\begin{equation}
    \label{eqn:g}
    g(x,y) = \left(N_1 +  N_2
    e^{-\frac{x^2+y^2}{2\sigma^2}}\right)\Theta(r_0^2-x^2-y^2),
\end{equation}
where the Gaussian is a consequence of the collimation, and the
unit step function $\Theta$ comes from the aperturing effect of
the Zeeman tube which has radius $r_0$, and $N_{1,2}$ are
normalization constants. Measurement of the OD is equivalent to
projecting (\ref{eqn:g}) along $x$, meaning we will be sensitive
to the $y$ mode function,
\begin{eqnarray}
    G(y) & \equiv & OD(y)/\int OD(y^\prime) \mathrm{d}y^\prime \\
    \label{eqn:G}
        & = & \int g(x,y)\mathrm{d}x.
\end{eqnarray}
Figs \ref{fig:normalabs}(b-d) show the data and corresponding fits
using (\ref{eqn:g},\ref{eqn:G}) for the atomic beam without
collimation (which has $N_2=0$), with 2D molasses, and with full
transverse collimation and cooling, respectively. Note that in all
three situations, the radius, $r_0$, of the disk is left as a fit
parameter, and yields a value near the actual 1.9 cm radius of the
Zeeman slower tube.

\begin{figure}
\includegraphics{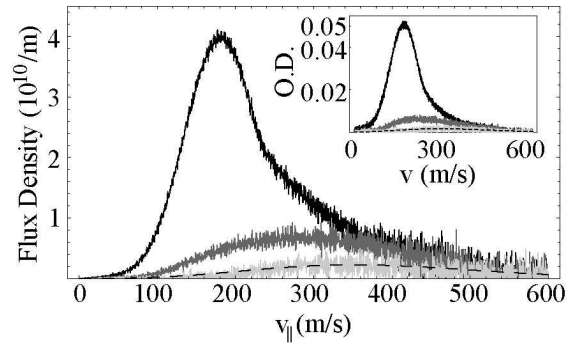}
\caption{\label{fig:9degrees} {\bf Flux enhancement through
transverse collimation and cooling}.  Flux measured  as a function
of longitudinal velocity based on measurement of the OD as a
function of probe frequency (inset). The color scheme here is as
in Fig.\,\ref{fig:normalabs}(a) The dashed curve is a calculation
of the distribution solely from effusion from the source at
$T=160^\circ\mathrm{C}$ (no adjustable parameters). }
\end{figure}

To measure the longitudinal velocity spread, we switch to a
geometry where the probe is at an angle $\phi = 9^\circ$ relative
to the normal of the atomic beam (Fig.\,\ref{fig:9degrees} inset).
The bottom curve shows the un-collimated effusive distribution
from the source. The middle and top curves represent the addition
of the 2D molasses and full transverse cooling and collimation,
respectively. As anticipated, both of these transverse cooling
configurations work best at lower longitudinal velocities, but
still give significant enhancement up to 300 m/s.

From this measurement, we may extract the flux by deconvolving the
Voigt distribution (\ref{eqn:voigt}) which may be written in terms
of the flux density, $\Phi(v) = n(v)/v$, at arbitrary probe angle
$\phi$ as
\begin{equation}
    \label{eqn:OD}
    OD(\Delta, y) = \frac{f\sigma_0 G(y)}{\cos\phi}\int_0^\infty \frac{\mathrm{d}v_\|}{v_\|}
    \frac{\Phi(v_\|)}{1+\frac{4}{\Gamma^2}(\Delta-k v_\| \sin\phi)^2}.
\end{equation}
Here, $v_\|$ is the longitudinal velocity, $\Delta$ is the
detuning of the probe from resonance, $\Gamma$ the natural
linewidth, and $\sigma_0$ the resonant cross section of the
transition which is weighted by the average oscillator strength
$f$.  (The probe is chosen to be linearly polarized, and averaging
over the available $F=2$ to $F=3$ transitions leads to $f \approx
0.46$.)

Without the Zeeman slower, in the limit where the width of the
longitudinal velocity distribution, contained in $\Phi(v_\|)$, is
much larger than $\Gamma$, we may pull the flux out of the $v_\|$
integral and rearrange to obtain
\begin{equation}
    \label{eqn:flux wide dist}
    \Phi(v_\|) = \frac{2 k v_\| \sin\phi \cos\phi}{f\sigma_0\Gamma\pi }
        \frac{OD(k v_\| \sin\phi, y=0)}{G(y=0)}.
\end{equation}
Thus, by measuring the OD at fixed $y=0$, having already mapped
out the beam's transverse mode to extract $G(y=0)$, we may readily
extract the atomic beam's flux density (Fig.\,\ref{fig:9degrees})
from the OD measurement (inset). The overall efficiency of the
transverse collimation and cooling process can be summarized as
follows: of the $1.1\times 10^{13}$/s which are ``capturable'' by
the Zeeman slower, only $2 \times 10^{11}$/s arrive in the main
chamber due to the solid angle of the Zeeman tube. Transverse
collimation and cooling increases this capturable flux to $5.3
\times 10^{12}$/s. This is a factor of 25 enhancement and means
that we can couple a full $50\%$ of the atoms which are usable
from our source through the Zeeman slower structure.

\begin{figure}
\includegraphics{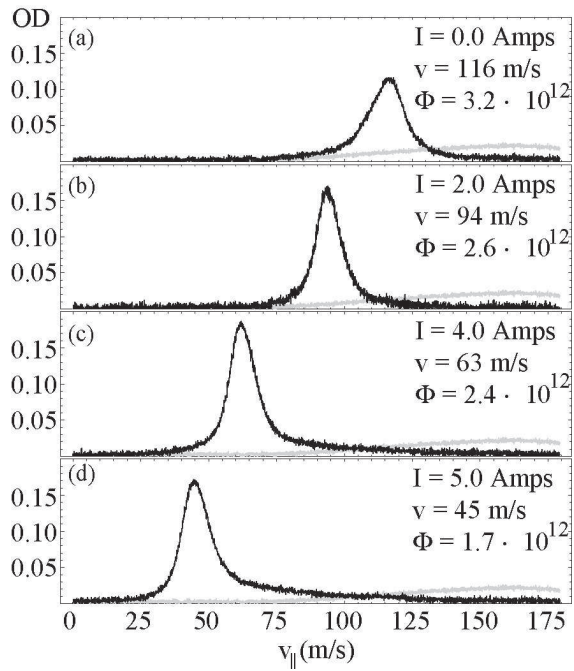}
\caption{\label{fig:Zeeman} {\bf Optical density measurement of
Zeeman slowed atomic beam} taken at a probe angle of $45^\circ$
relative to the atomic beam axis for varying final velocities.  In
all cases, the top thirteen coils of the slower are maintained at
$I=5$ A (corresponding to a capture velocity of 320 m/s).  The
current in each plot is that of the final set of Zeeman slower
coils, which adjust the final velocity by 16 (m/s)/A. The grey
curve represents the same situation as the black curve in
Fig.\,\ref{fig:9degrees}. }
\end{figure}

The measured effect of the Zeeman slower on the atomic beam with
various magnetic field configurations is shown in
Fig.\,\ref{fig:Zeeman}. In this series, we choose our probe beam
to be at $45^\circ$ to the axis of the atomic beam, to allow for
maximum velocity sensitivity.  In Fig.\,\ref{fig:Zeeman}(a) we
show the effect of turning on the positive field of the Zeeman
slower (as defined by the first thirteen coils in
Fig.\,\ref{fig:setup}(d)). As we increase the field in the
negative region as shown in Figs \ref{fig:Zeeman}(b-d), we can
continuously tune the final velocity of our atomic beam, while
always maintaining a high overall flux (for the cases shown,
tuning the velocity from 116 to 45 m/s).

With the slower activated, the width of the longitudinal velocity
distribution is narrow, and we may substitute $1/v_0$ for $1/v_\|$
in (\ref{eqn:OD}), where $v_0$ is the average velocity of the
atomic beam.  We integrate and rearrange to get the total flux
\begin{equation}
    \label{eqn:total flux}
    \int \Phi(v_\|) \mathrm{d}v_\| = \frac{2 v_0 \cos\phi}{f\sigma_0\Gamma\pi G(y)
    } \int OD(\Delta, y)\mathrm{d}\Delta.
\end{equation}
So, by measuring the spatial mode, $G(y)$ of the atomic beam in
each of the cases in Fig. \ref{fig:Zeeman} (as we did in Figs
\ref{fig:normalabs}(b-d)), we may readily compute the total flux
of the atomic beam.

Note that in these figures, the stated flux results from
integrating the measured OD over the peak. As we proceed to lower
final velocities, the flux in this peak gradually drops and the
distribution begins to form a tail in the high velocity region. In
fact, integrating over the total distribution (up to 150 m/s)
yields a total flux of $3\times 10^{12}$ atoms/s in all cases, and
the presence of the high tail has been traced to minor ripples in
the magnetic field at the end of the Zeeman slower.

Importantly, the signal enhancement for the Zeeman slowed
distribution due to the transverse collimation and cooling exceeds
the factor of 25 observed for the unslowed beam. We measure
$3.2\times 10^{12}$ ($1.7\times 10^{12}$) atoms/s at 116 (45) m/s,
to be compared to $6\times 10^{10}$ ($3\times 10^{10}$) atoms/s
without any transverse collimation and cooling, for factor of 50
in overall increase.

\begin{acknowledgments}

This work was supported by the Air Force Office of Scientific
Research, ARO-MURI, and NASA.  CS was supported by a National
Defense Science and Engineering Grant sponsored by the U.S.
Department of Defense.  The authors would like to acknowledge D.
Rogers, J. MacArthur and A. Sliski for invaluable technical
expertise and thank W. H\"ansel for useful discussions and
valuable support.

\end{acknowledgments}


\end{document}